# On the theory of time dilation in chemical kinetics[*]

Mirza Wasif Baig[1]

[1] *J. Heyrovský Institute of Physical Chemistry, Academy of Sciences of the Czech Republic, CZ-18223 Prague 8, Czech Republic*

*wasifbaig.mirza@jh-inst.cas.cz*

The rates of chemical reactions are not absolute but their magnitude depends upon the relative speeds of the moving observers. This has been proved by unifying basic theories of chemical kinetics, which are transition state theory, collision theory, RRKM and Marcus theory with the special theory of relativity. Boltzmann constant and energy spacing between permitted quantum levels of molecules are quantum mechanically proved to be Lorentz variant. The relativistic statistical thermodynamics has been developed to explain quasi-equilibrium existing between reactants and activated complex. The newly formulated Lorentz transformation of the rate constant from Arrhenius Equation, of the collision frequency and of the Eyring and Marcus equations renders the rate of reaction to be Lorentz variant. For a moving observer moving at fractions of the speed of light along the reaction coordinate the transition state possess less kinetic energy to sweep translation over it. This results in the slower transformation of reactants into products and in a stretched time frame for the chemical reaction to complete. Lorentz transformation of the half-life equation explains time dilation of the half life period of chemical reactions and proves special theory of relativity and presents theory in accord with each other. To demonstrate the effectiveness of the present theory, the enzymatic reaction of methylamine dehydrogenase and radioactive disintegration of Astatine into Bismuth are considered as numerical examples.

## 1. Introduction

The special theory of relativity proposed in 1905 by Einstein [1,2] is based on two basic postulates i.e. (i) *Speed of light is a kind of cosmic speed limit,* (ii) *All laws of physics hold good in all inertial frames.* It explains different physical quantities i.e. time, length and mass observe to change for observers moving at fractions of speed of light in terms of Lorentz Transformations. It explains that the time lapse between two events is not constant from one observer to another, rather it is dependent on the relative speeds of the observer's reference frames, this phenomenon is known as time dilation and is mathematically expressed by the following Lorentz transformation, [3]

$$t_u = \gamma t_o \tag{1}$$

Where Lorentz transform factor in Eq. (1) is "$\gamma = (1 - u^2/c^2)^{-1/2}$". In present paper all variables with subscript *o* refer to the inertial frame $K_o$ with the observer at rest, while variables with subscript *u* refer to inertial frame $K_u$ with the observer moving at relative speed *u*. The Lorentz transformation for mass has been experimentally verified for ions moving at high speed in heavy ion accelerators.[4] The Lorentz transformation for mass is back bone of relativistic quantum chemistry.[5] The corresponding equations are as follows,

$$m_u = \gamma m_o \tag{2}$$

$$L_u = \gamma^{-1} L_o \tag{3}$$

## 2. Theory

The present study is the first theoretical attempt to explain time dilation in chemical reactions. The phenomenon of time dilation has been experimentally verified for fast moving muons. [6,7] More than two decades ago Ohsumi published an article based on relativistic motion while discussing reaction kinetics.[8] The author stated invariance of the reaction rate and covariance of the rate constant under Lorentz transformations. This work fails to explain time dilation for the half life of chemical and nuclear reactions and how the reaction slows down and prolongs for longer time in moving frames. However, the major drawback in the paper is that it supports a temperature transformation i.e. $T_u = \gamma^{-1} T_o$ [9] and contradicts with $T_u = \gamma T_O$.[10] If either of these temperature transformations had been possible, it would have drastically disturbed the chemical system at equilibrium. However, both of these temperature transformations are wrong, and as recently has been proved by Landsberg and Matsas considering Unruh-Dewitt detector that $T_u = T_O = T$ and no universal temperature transformation exists. [11] Neither of any temperature transformation can explain time dilation in nuclear and chemical reactions. These findings outdate the relativistic kinetics given by Ohsumi.[8] To explain time dilation firstly we develop relativistic statistical mechanics and relativistic thermodynamics for quasi equilibrium existing between reactants and activated complexes.

## 3. Relativistic Statistical Mechanics

From spectroscopic studies of molecules it is found in nature that electronic transitions are very much rapid than vibrational transition of molecules. Vibrational transitions are also found to be faster than rotational transitions. [12-14] This is because electronic energy levels are widely separated than vibrational energy levels and vibrational energy levels are widely separated than rotational energy levels. Even to attain population inversion in lasers of three level systems, energy spacing between ground and second energy level is kept greater than energy spacing between second and third energy level. [15] This is because smaller the spacing between two levels of system, longer the system can remain in excited energy level. According to Heisenberg time energy uncertainty relation gives the uncertainty of the energy of a state with life time as, [16-17]

$$\langle \Delta \varepsilon_o \rangle \langle \Delta t_o \rangle \geq \hbar \qquad (4)$$

It follows that energy of the state $\langle \Delta \varepsilon_o \rangle$ would be exactly defined, and the state would be truly stationary state, only if the life time $\langle \Delta t_o \rangle$ were infinite. In reality, time $\langle \Delta t_o \rangle$ is not infinite and the state is represented by energies smeared over a range. For moving observer Heisenberg time energy uncertainty relation can be written as,

$$\langle \Delta \varepsilon_u \rangle \langle \Delta t_u \rangle \geq \hbar \qquad (5)$$

To be consistent with time dilation, life time of an energy state will also be stretched i.e. $t_u = \gamma t_0$. Placing this Lorentz transformation for life time of energy state in Eq. (2) gives,

$$\langle \Delta \varepsilon_u \rangle \gamma^{-1} \langle \Delta t_u \rangle \geq \hbar \qquad (6)$$

For Heisenberg time energy uncertainty to remain invariant for moving observer, energy of state smeared over a range $\langle \Delta \varepsilon_u \rangle$, should contract and become Lorentz variant i.e. $\langle \Delta \varepsilon_u \rangle = \gamma^{-1} \langle \Delta \varepsilon_o \rangle$. Placing this Lorentz transformation for energy levels in Eq. (6) mathematically proves Heisenberg time energy uncertainty to be Lorentz invariant and retains it's from in all inertial frames i.e.

$$\gamma \langle \Delta \varepsilon_u \rangle \gamma^{-1} \langle \Delta t_u \rangle \geq \hbar \qquad (7)$$

Time dilation which is an experimentally verified fact, [6-7] dictates energy of permitted energy levels and spacing between then to be Lorentz variant i.e. $\langle \Delta \varepsilon_u \rangle = \gamma^{-1} \langle \Delta \varepsilon_o \rangle$. From spectroscopic studies of molecules it is found that following inequalities hold for electronic, vibrational and rotational transitions at room temperature i.e. [16]

$$\langle \Delta \varepsilon_{elec} \rangle_o \ggg (k_B)_0 T \qquad (8)$$

$$\langle \Delta \varepsilon_{vib} \rangle_o > (k_B)_0 T \qquad (9)$$

$$\langle \Delta \varepsilon_{rot} \rangle_o \simeq (k_B)_0 T \qquad (10)$$

Inequalities of Eq. s. (8), (9) and (10) for electronic, vibrational and rotational transitions of molecules at room temperature should also hold for moving observer i.e.

$$\langle \Delta \varepsilon_{elec} \rangle_u \ggg (k_B)_u T \qquad (11)$$

$$\langle \Delta \varepsilon_{vib} \rangle_u > (k_B)_u T \qquad (12)$$

$$\langle \Delta \varepsilon_{rot} \rangle_u \simeq (k_B)_u T \qquad (13)$$

As time dilation dictates spacing between allowed energy levels to be Lorentz variant i.e. $\langle \Delta \varepsilon_u \rangle = \gamma^{-1} \langle \Delta \varepsilon_o \rangle$. So for inequalities given in Eq. s. (11), (12) and (13) to remain valid for moving observer either Boltzmann constant or temperature has to be Lorentz variant. As recently Landsberg and Matsas considering Unruh-Dewitt detector proved that $T_u = T_o = T$ and no universal temperature transformation exists.[11] Moreover for an equilibrium thermodynamic system at a triple point of the corresponding *P-T* diagram with respect to observer $K_o$, three phases will coexist. For instance, these could be graphite, diamond, and liquid carbon (at extremely high pressure and temperature) or it could be the triple point of ice and water, etc. The same three phases will be seen from any other moving observer $K_u$. As soon as equilibrium thermodynamic system can coexist at one *P, T* point only, the temperature and the pressure of this system will be the same for all observers irrespective of their speeds and thus proving them Lorentz invariant.[18] Time dilation at molecular level is only possible if spacing between permitted energy states decrease for moving observer. As spacing between permitted energy states is independent of temperature and cannot be altered by either increasing or decreasing temperature. Since neither of any possible Lorentz transformations of temperature [8-9] can support Lorentz transformation of spacing between permitted energy levels, thus no universal Lorentz transformation of temperature exists. Therefore Lorentz transformation of spacing between permitted energy levels dictates Boltzmann constant to be Lorentz variant i.e. $(k_B)_u = \gamma^{-1}(k_B)_0$ and temperature to be Lorentz invariant. Lorentz transformation of Boltzmann constant and ideal gas constant has already reported. [18-19]

### 3.1. Relativistic Maxwell Boltzmann Distribution Law

Avramov proposed Boltzmann constant to be Lorentz variant; but with that he proposed statistical physics to suffer with serious consequences.[18] So Avramov's view that Lorentz transformation of Boltzmann constant is a tempting problem for statistical thermodynamics contradicts with second postulate of special relativity which states laws of physics retain their form in all inertial frames. According to Maxwell Boltzmann Distribution law number of material particles i.e. molecules in *jth* energy level $\varepsilon_i$ for moving observer is given by following relation, [20]

$$\langle n_j \rangle_u = exp\left(\frac{-\langle \Delta \varepsilon_j \rangle_u}{(k_B)_u T}\right) \tag{14}$$

Substituting Lorentz transformation of spacing between quantized energy level $\langle \Delta \varepsilon_u \rangle = \gamma^{-1} \langle \Delta \varepsilon_o \rangle$ and Boltzmann constant $(k_B)_u = \gamma^{-1}(k_B)_0$ respectively in Eq. (14) gives Maxwell Boltzmann distribution of molecules to be Lorentz invariant and number of molecules in energy level is same irrespective of relative speed of observers.

$$\langle n_j \rangle_u = \langle n_j \rangle_o = n_j \tag{15}$$

### 3.2. Relativistic Molecular Partition Function

Maxwell Boltzmann statistics explains distribution of weakly coupled distinguishable material particles i.e. atoms and molecules over various energy states in thermal equilibrium, when the temperature is high enough and density is low enough and quantum effects are negligible. [20-21] In Maxwell Boltzmann Statistics total partition function is product of translational, rotational, vibrational and electronic partition function, however at room temperature electronic partition function does not contribute towards thermodynamic state variables. Total partition function in terms of translational, rotational, vibrational and electronic partition functions for moving observer partition function can be defined as,

$$(Q_{total})_u = \frac{1}{N!}(q^T)_u^N (q^R)_u^N (q^V)_u^N (q^E)_u^N \tag{16}$$

### 3.2.1. Relativistic Translational Partition Function

Translational partition functions for weakly interacting molecules of mass $m_u$ confined in volume $V$ is defined in Eq. (17). [20-21] Here Lorentz transformation of volume will not be considered because this can be explained by considering symmetric box like cubic box there are many degenerate energy states exists due to symmetry so if Lorentz transformation of volume is taken in to account it will uplift the degeneracy among the different energy states so Lorentz transformation of volume is discarded here. Energy of all degenerate energy levels should be equally lowered for moving observers equally in order to achieve symmetry conservation in moving frame and thus Lorentz transformation of volume is being neglected here. So translational partition function for moving observer will acquire following form,

$$(q^T)_u = (2\pi m_u (k_B)_u T/h^2)^{(3/2)} V \qquad (17)$$

Substituting Lorentz transformations of Boltzmann constant and mass in Eq. (17) proves translational partition function Lorentz invariant i.e.

$$(q^T)_u = (q^T)_o = q^T \qquad (18)$$

### 3.2.2. Relativistic Rotational Temperature and Partition Function

Rotational partition function for diatomic molecules in terms of rotational temperature is defined as, [20-21]

$$(\Theta_R)_u = hcB_u/(k_B)_u \qquad (19)$$

For moving observer rotational constant is defined as,

$$B_u = h / 8\pi^2 c I_u \qquad (20)$$

For moving observer mass of electron increase so according to Heisenberg's Uncertainty principle i.e. "$m\Delta v \Delta x \geq \hbar$" so velocity of electron revolving should slows down for moving observer while space occupied by electrons remains invariant i.e. why bond length is Lorentz invariant. Since mass is Lorentz variant $m_u = \gamma m_o$ so moment of inertia becomes Lorentz variant i.e. $I_u = \gamma I_o$. Substituting Lorentz transformation in Eq. (20) makes rotational constant Lorentz variant as,

$$B_u = \gamma^{-1} B_o \qquad (21)$$

Placing Lorentz transformation of Boltzmann constant and rotational constant $(k_B)_u = \gamma^{-1}(k_B)_0$ and $B_u = \gamma^{-1} B_o$ respectively in Eq. (21) renders rotational temperature Lorentz invariant,

$$(\Theta_R)_u = (\Theta_R)_o = \Theta_R \qquad (22)$$

Eq. (22) proves that like translational temperature rotational temperature is also Lorentz invariant. This is first time rotational temperature is reported to be Lorentz variant. Rotational partition function for moving observer is defined as,

$$(q^R)_u = T/\sigma \, (\Theta_R)_u \qquad (23)$$

As from Eq. (22) rotational temperature is Lorentz invariant i.e. $(\Theta_R)_u = (\Theta_R)_o = \Theta_R$, so this renders rotational partition function also to be Lorentz invariant,

$$(q^R)_u = (q^R)_o = q^R \tag{24}$$

### 3.2.3. Relativistic Vibrational Temperature and Partition Function

Vibrational partition function of molecules for moving observer in terms of vibrational temperature is defined as, [20-21]

$$(q^V)_u = 1/1 - ((\Theta_{vib})_u/T) \tag{25}$$

Vibrational temperature in terms of frequency of harmonic oscillator $v_u$ for moving observer is defined as,

$$(\Theta_{vib})_u = hv_u/(k_B)_u \tag{26}$$

Where $v_u$ is frequency of harmonic oscillator for moving observer, as for moving observer mass of atoms of molecules increases so frequency of vibrating atoms decreases and hence it becomes Lorentz variant $v_u = \gamma^{-1}v_o$. Decrease in frequency of vibrating atoms for moving observer is in accord with phenomenon of time dilation. Placing Lorentz transformations of Boltzmann constant and frequency of oscillating atom in Eq. (26) proves vibrational temperature Lorentz invariant.

$$(\Theta_{vib})_u = (\Theta_{vib})_o = \Theta_{vib} \tag{27}$$

Eq. (27) proves that like translational temperature vibrational temperature is also Lorentz invariant. This is again first time vibrational temperature is reported to be Lorentz variant.

As from Eq. (27) vibrational temperature is Lorentz invariant i.e. $(\Theta_{vib})_u = (\Theta_{vib})_o = \Theta_{vib}$, so this renders vibrational partition function also to be becomes Lorentz invariant,

$$(q^V)_u = (q^V)_o = q^V \tag{28}$$

### 3.2.4. Electronic Temperature and Partition Function

Electronic partition function for atoms and molecules in terms of electronic temperature for moving observer is defined as, [20-21]

$$(q^E)_u = \sum_i^N g_i exp(-(\Theta_E)_u/T) \tag{29}$$

Electronic temperature for moving observer is defined as,

$$(\Theta_E)_u = h(\varepsilon_i^E)_u/(k_B)_u \tag{30}$$

As it follows from Eq. (7) that energy of *jth* electronic level is Lorentz variant i.e. $(\varepsilon_i^E)_u = \gamma^{-1}(\varepsilon_i^E)_o$. Placing Lorentz transformations of Boltzmann constant and energy of *jth* electronic level in Eq. (30) renders electronic temperature Lorentz invariant i.e.

$$(\Theta_E)_u = (\Theta_E)_o = \Theta_E \tag{31}$$

From Eq.s (31) electronic partition function can be concluded as Lorentz invariant.

$$(q^E)_u = (q^E)_o = q^E \tag{32}$$

From Eq.s (18), (24), (28) and (32) total molecular partition function is found to be Lorentz invariant and its value is independent of relative speed of observers.

$$(Q_{total})_u = (Q_{total})_o = Q_{total} = Q \tag{33}$$

### 3.3. Relativistic Statistical Thermodynamics

Statistical thermodynamics is language of expressing all thermodynamic state functions in terms of molecular partition function. Statistical mechanics provides a molecular-level interpretation of macroscopic thermodynamic quantities such as work, free energy and entropy. It enables the thermodynamic properties of bulk materials to be related to the spectroscopic data of individual molecules. In statistical mechanics partition function $Q$ encodes all properties of system in thermodynamics equilibrium. Behavior of all thermodynamic state variables and equilibrium constant is very well explained in terms of partition function. [20] For moving observer they all can be expressed as,

$$U_u = -N(k_B)_u (\partial Q/\partial T)_V \tag{34}$$

$$H_u = T[(k_B)_u T(\partial \ln Q/\partial T)_V + V_u (k_B)_u (\partial \ln Q/\partial V_u)_T] \tag{35}$$

$$S_u = (k_B)_u T(\partial \ln Q/\partial T) + (k_B)_u \ln Q \tag{36}$$

$$A_u = (k_B)_u T \ln Q \tag{37}$$

$$G_u = (k_B)_u T[(\ln Q) - V_u (\partial \ln Q/\partial V_u)_T] \tag{38}$$

Substituting Lorentz transformations of Boltzmann constant and volume in Eq.s (34), (35), (36), (37) and (38) gives following Lorentz transformations for all thermodynamic state functions.

$$U_u = \gamma^{-1} U_o \tag{39}$$

$$H_u = \gamma^{-1} H_o \tag{40}$$

$$S_u = \gamma^{-1} S_o \tag{41}$$

$$A_u = \gamma^{-1} A_o \tag{42}$$

$$G_u = \gamma^{-1} G_o \tag{43}$$

Lorentz transformations of all thermodynamic state functions are in accord with Lorentz transformation of thermodynamic work. These Lorentz transformations of thermodynamic state functions can explain quasi equilibrium existing between reactants and activated complexes during course of reaction.

## 4. Relativistic Chemical Kinetics

### 4.1. Theories of Chemical kinetics for Moving Observer

To theoretically explain the stretching of the time frame of chemical reactions for an observer moving at fractions of the speed of light along the reaction coordinate, the rates of reactions should slow down. Since the speed of chemical reactions is quantitatively explained through basic theories of chemical kinetics, these theories have to be merged with the special theory of relativity in order to explain time dilation for chemical reactions. The necessary mathematical forms of the rate laws in three basic theories of chemical kinetics, meeting the requirements of special relativity, are derived in the following. It is shown that the necessary requirements can be met for the transition state theory of chemical reactions, for the collision theory for bimolecular reactions, and for the Marcus theory of electron transfer reactions, if the respective rate constants are no longer considered as Lorentz invariants but are allowed to become Lorentz variants.

### 4.1.1. Transition State Theory of Chemical Reactions for Moving Observer

Transition state theory introduced by Eyring, Evans and Polanyi in 1935 [22-23] separating reactants and products on potential energy surface is used to formulate an expression for thermal rate constant, which has been derived by assuming that electronic and nuclear motions are separate which is equivalent to Born-Oppenheimer approximation. Reactant molecules are distributed among their states in accordance with Maxwell Boltzmann Distribution. Even in the absence of equilibrium between reactant and product molecules, the transition states that are becoming products are distributed among their states according to Maxwell Boltzmann Distribution laws. There exist quasi-equilibrium between reactants and activated complexes. In the transition state motion along the reaction coordinate is assumed to be separated from the other motions and treated classically as translational. Marcus while unifying RRK theory with transition state theory also treats motion of transition state along the reaction coordinate as a simple translation.[24] For thermal averaging in transition state theory to be meaningful, it is necessary that translational energy of transition sate in reaction coordinate must be less than $(k_B)_o T$. [25] According to time energy Heisenberg uncertainty principle's $(E)_o (\Delta t)_o \geq \hbar$, life time of transition state "$(\Delta t)_o$" must be larger than "$\hbar/(k_B)_o T$". Mathematically it can be written as,

$$(k_B)_o T (\Delta t)_o \geq \hbar \tag{44}$$

For moving observer Heisenberg uncertainty principle becomes,

$$(k_B)_u T (\Delta t)_u \geq \hbar \tag{45}$$

Substituting Lorentz transformation of time in Eq. (45) gives;

$$(k_B)_u T \gamma (\Delta t)_o \geq \hbar \tag{46}$$

As special theory of relativity states that laws of physics remain same in all inertial frames same, thus Heisenberg uncertainty principle should also restore its form i.e.

$$\gamma^{-1} (k_B)_u T \gamma (\Delta t)_o \geq \hbar \tag{47}$$

As temperature is Lorentz invariant $T_u = T_o = T$ [11] so this quantum mechanically proves Boltzmann constant Lorentz variant, this is also proposed by I. Avramov, [12]

$$(k_B)_u = \gamma^{-1}(k_B)_0 \tag{48}$$

Since Avogadro number is independent of the relative speeds of moving observers, therefore same transformation can be written for ideal gas constant i.e. [14]

$$R_u = \gamma^{-1} R_0 \tag{49}$$

Eyring equation reformulates rate constant for $n^{th}$ order reaction in thermodynamic terms, which transforms for moving as; [25-26]

$$(k_n)_u = \kappa (k_B)_u T/h [(c^0)]^{(n-1)} exp\left((\Delta S^\dagger)_u/R_u\right) exp\left((-\Delta H^\dagger)_u/R_u T\right) \tag{50}$$

Substituting Lorentz transformations of enthalpy, entropy and Boltzmann constant from Eq. (40) (41) and (48) respectively in Eq. (50) gives,

$$(k_n)_u = \gamma^{-1}(k_n)_0 \tag{51}$$

Arrhenius factor quantitatively explains number of reactant molecules crossing energy barrier and transforming into products in thermodynamic terms. It transforms for moving observer as; [25]

$$A_u = exp[-(\Delta n - 1)](k_B)_u T/h[(c^0)]^{(n-1)} exp\left((\Delta S^\dagger)_u/R_u\right) \tag{52}$$

Placing Lorentz transformations of the Boltzmann constant, the universal gas constant $R_u = \gamma^{-1} R_0$ and entropy $(\Delta S^\dagger)_u = \gamma^{-1}(\Delta S^\dagger)_o$ in Eq. (54) proves Arrhenius factor Lorentz variant i.e.

$$A_u = \gamma^{-1} A_0 \tag{53}$$

Arrhenius factor which is a frequency and its decrement in magnitude turns out to be totally in agreement with time dilation phenomenon. Arrhenius equation explains rate constant as an exponential function of activation energy which transforms for moving observer as,

$$(k_n)_u = A_u exp((-E_a)_u/R_u T) \tag{54}$$

Like all other thermodynamic state functions Activation energy should also be Lorentz variant i.e $(E_a)_u = \gamma^{-1}(E_a)_0$. Inserting Lorentz transformations of Activation energy, Arrhenius factor and the universal gas constant, $(E_a)_u = \gamma^{-1}(E_a)_0$, $A_u = \gamma^{-1} A_0$ and $R_u = \gamma^{-1} R_0$ respectively in Eq. (54) proves rate constant Lorentz variant,

$$(k_n)_u = \gamma^{-1}(k_n)_0 \tag{55}$$

Eyring equation for reaction $A + B \rightarrow AB^\ddagger \rightarrow P$ in terms of partition function for moving observer can be stated as, [26]

$$(k_n)_u = \kappa (k_B)_u T/h \left(Q_{AB}^\dagger/Q_A Q_B\right) exp(-(E_a)_u/R_u T) \tag{56}$$

As from Eq. (33) molecular partition function is Lorentz invariant and by subsituting value of Lorentz transformations of Activation energy, the Boltzmann constant and the universal gas constant in Eq. (56)

proves rate constant to be Lorentz variant i.e. $(k_n)_u = \gamma^{-1}(k_n)_0$. Similarly from RRKM theory which generalizes to transition state theory at high pressure limit rate constant of unimolecular reactions defined as $(k_n^\infty)_u = (k_B)_u T/h \left(Q_r^\dagger Q_v^\dagger/Q_r Q_v\right) exp(-(E_0)_u/R_u T)$, gives the same Lorentz transformation for the rate constant $(k_n^\infty)_u = \gamma^{-1}(k_n^\infty)_0$.[25]

### 4.1.2. Collision Theory of Bimolecular Reactions for Moving Observer

According to the collision theory for reaction rates, the molecules of reactants are considered as hard sphere colliding with each other with the assumption that rate of reaction depends upon number of collisions. The theory suggests rate of reaction in terms of important parameters (i) collision frequency, (ii) collision cross section and (iii) relative velocity as the.[25-27] Mostly collision theory gives best explanation of bimolecular reactions like A + B → P. Collision theory expresses rate constant in terms of collision cross section and relative velocity of colliding molecules for moving observer which transform as,

$$(k_2)_u = N_A \sigma_{AB} (v_r)_u \tag{57}$$

Collision cross section area is Lorentz invariant as structure of colliding atoms and molecules remain unchanged. Relative velocity between the colliding atoms and molecules transforms for moving observer as,

$$(v_r)_u = (8(k_B)_u T/\pi \mu_u)^{1/2} \tag{58}$$

Therefore, placing Lorentz transformations of the Boltzmann constant and mass in Eq. (58) proves relative velocity of the colliding molecules Lorentz variant.

$$(v_r)_u = \gamma^{-1}(v_r)_u \tag{59}$$

Relative velocity between the colliding atoms and molecules decreases for moving observer as Boltzmann constant decreases so tendency of molecules to execute translational motion slows down. Placing value of relative velocity $(v_r)_u = \gamma^{-1}(v_r)_u$ from Eq. (59) in Eq. (57) again proves rate constant Lorentz variant and it decreases for moving observer,

$$(k_2)_u = \gamma^{-1}(k_2)_0 \tag{60}$$

Collision frequency for bimolecular in terms of mole densities, $\rho_A \rho_B$ collision cross section and relative velocity for moving observer can be stated as,

$$(Z_{AB})_u = N_A \sigma_{AB} (v_r)_u \rho_A \rho_B \tag{61}$$

Placing Lorentz transformations of relative velocity Eq. (59) in Eq. (61) shows that collision frequency decreases for moving observer,

$$(Z_{AB})_u = \gamma^{-1}(Z_{AB})_o \tag{62}$$

Collision frequency being reciprocal of time decreases for moving observer which is in accordance with time dilation.

### 4.1.3. Marcus Theory of Electron Transfer Reactions for Moving Observer

A theoretical model for electron transfer reactions especially for outer sphere electron transfer reactions was developed by Marcus. [28-32] This model envisages the solvent around the reactant ions first configured to be favorable for electron transfer. There is a solvent configuration around each reactant ion for which the Gibb's free energy $G$ is a minimum and changes in the solvent structure from this configuration increases the free energy. To attain the transition state, for successful electron transfer the separation between the two reactant ions decreases and reorganization of the solvent structure about each ion occurs. A reaction coordinate for electron transfer may be conceived as a combination of these ion-ion separations and solvent reorganization coordinates. Gibb's free energy of reactants and products versus reaction coordinate is a parabolic function. Transition state is located at a point, where two parabolic curves of reactants and products intersect each other. Marcus provides a formula for the activation energy based on a parameter called the reorganization energy or Gibbs free energy. The Marcus expression for rate constant of a pure electron transfer reaction $A^{ZA} + B^{ZB} \rightarrow A^{ZA + \Delta Z} + B^{ZB -- \Delta Z}$ for moving observer will transform as,

$$(k_{AB})_u = (Z_{AB})_u \exp\left(-\left((\Delta G^0_{AB})_u + \lambda_u\right)^2 / 4\lambda_u R_u T\right) \tag{63}$$

$\lambda_u$ is reorganization energy, it is defined as the energy required to "reorganize" the system structure from initial to final coordinates, without changing the electronic state. Like other thermodynamic parameters, reorganization energy is Lorentz variant and it can be proved as follows. Reorganization energy is composed of vibrational and solvational$(\lambda_i)_u$ and $(\lambda_0)_u$ components respectively.

Vibrational reorganization energy $(\lambda_i)_u$ is expressed in terms of reduced force constant $(k_j)_u$ of the $j^{th}$ normal mode coordinates of reactants $q_j^r$ and products $q_j^p$.

$$(\lambda_i)_u = 1/2 \sum_j (k_j)_u \left(q_j^r - q_j^p\right)^2 \tag{64}$$

Reduced force constant $(k_j)_u$ of the $j^{th}$ normal mode is given by $(k_{AB})_u = 4\pi^2 \omega_u^2 \mu_u^2$. Substituting Lorentz transform equations of reduced mass $\mu_u = \gamma \mu_o$ and oscillation frequency $\omega_u = \gamma^{-1} \omega_o$ proves force constant to be Lorentz variant as $(k_j)_u = \gamma^{-1}(k_j)_0$. On substituting Lorentz transformed equation for rate constant in Eq. (68) proves vibrational reorganization energy Lorentz variant as,

$$(\lambda_i)_u = \gamma^{-1}(\lambda_i)_0 \tag{65}$$

For $\Delta e$ charge transferred between reactants solvational reorganization energy is mathematically expressed in terms of ionic radii $a_1$ and $a_2$ and, the centre to centre separation distance of the reactants $W$, refractive index and dielectric constants of the solvent which are $n_u$ and $\varepsilon_u$ respectively.

$$(\lambda_0)_u = (\Delta e)^2 (1/2a_1 + 1/2a_2 - 1/W)(1/(n_s)_u^2 - 1/(\varepsilon_s)_u) \tag{66}$$

Since increase in mass renders density Lorentz variant i.e. and this would make dielectric constant and refractive index of solvent and becomes Lorentz variant i.e. $(\varepsilon_s)_u = \gamma(\varepsilon_s)_0$ and $(n_s)_u = \gamma^{1/2}(n_s)_0$ respectively. Placing values of Lorentz transformed equations of refractive index dielectric constant in Eq. (66) gives following Lorentz transformation for solvational reorganization energy,

$$(\lambda_0)_u = \gamma^{-1}(\lambda_0)_o \tag{67}$$

Therefore, from Eq. (65) and (67) total reorganization energy becomes Lorentz variant i.e. $\lambda_u = \gamma^{-1}\lambda_o$. Substituting values of Lorentz transformed equations for free energy $(G_{AB})_u = \gamma^{-1}(G_{AB})_o$, reorganization energy $\lambda_u = \gamma^{-1}\lambda_o$, collision frequency $(Z_{AB})_u = \gamma^{-1}(Z_{AB})_o$ and ideal gas constant $R_u = \gamma^{-1}R_o$ respectively in Eq. (64) proves rate constant for electron transfer reactions Lorentz variant,

$$(k_{AB})_u = \gamma^{-1}(k_{AB})_o \tag{68}$$

Let an electron transfer reaction in which A, B are reactants and $X^*$, X are hypothetical initial and final thermodynamic states of the system called intermediates.

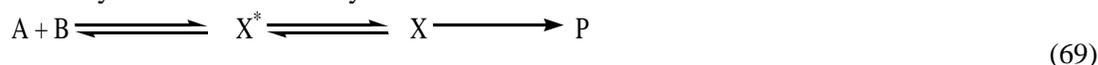

$$\tag{69}$$

When reactants are near each other suitable solvent fluctuation result in formation of state $X^*$, whose atomic configuration of the reacting pair and of the solvent is that of the activated complex, and whose electronic configuration is that of the reactant. $X^*$ can either form the reactant following disorganization of some of the oriented solvent molecules, or it can form state X by an electronic transition, X has atomic configuration which is same as that of $X^*$ but it has an electronic configuration which is that of products. The pair of states $X^*$ and X constitute activated complex. Marcus and Sutin have formulated rate constant for electron transfer reaction in terms of electronic coupling $(H_{AB})_o$ between the initial and final state of the electron transfer reaction (i.e., the overlap of the electronic wave functions of the two states). For moving observer rate constant can be written as, [29]

$$(k_{et})_u = 2\pi/\hbar \, (H_{AB})_u^2 \left(1/\sqrt{4\pi\lambda_u R_u T}\right) exp(-((\Delta G^0)_u + \lambda_u)^2/4\lambda_u R_u T) \tag{70}$$

Substituting Lorentz transformations for free energy i.e. $(\Delta G^0)_u = \gamma^{-1}(\Delta G^0)_o$, electronic coupling $(H_{AB})_u = \gamma^{-1}(H_{AB})_o$, universal gas constant $R_u = \gamma^{-1}R_o$ and reorganization energy $\lambda_u = \gamma^{-1}\lambda_o$ gives Lorentz transformed equation for electron transfer reaction as,

$$(k_{et})_u = \gamma^{-1}(k_{et})_o \tag{71}$$

## 5. Lorentz transformation of rate of reaction

From the knowledge of chemical kinetics, it is known that the rate of a chemical reaction is defined as the rate of change of concentration "$C$" with respect to time t. [28-32] In case of gas phase reaction "$C$" is replaced by pressure "$P$" and number of molecules or atoms "$N$" in solid phase reactions (nuclear reactions), all of these three quantities are Lorentz invariant. For moving observer the rate law can be stated as,

$$(r_n)_u = d[C]/dt_u = (k_n)_u[C]^n \tag{72}$$

Substituting Lorentz transformations of rate constant $(k_n)_u = \gamma^{-1}(k)_o$ in Eq. (72) proves rate of reaction Lorentz variant.

$$(r_n)_u = \gamma^{-1}(r_o)_o \tag{73}$$

## 6. Lorentz transformation of half Life

Half-life period is the time period during which initial concentration $C_o$ of a reactant reduces to one half of its initial value. Equation for half life period of reaction for moving observer can be written as, [25-27]

$$(t_{1/2})_u = J/(k_n)_u [C_o]^{(n-1)} \qquad (74)$$

$J$ is the coefficient for $n^{th}$ order chemical reaction. Substituting Lorentz transformations of rate constant $(k_n)_u = \gamma^{-1}(k)_o$ in Eq. (74) gives Lorentz transformation for half- life,

$$(t_{1/2})_u = \gamma (t_{1/2})_o \qquad (75)$$

Lorentz transformed equation for half life is similar to Einstein's time dilation equation, where former explains time dilation in the chemical (molecular) world while the latter explains time dilation in the physical world.

## 7. Relativistic Equilibrium Constant

Consider following chemical reaction at chemical equilibrium, [20-21]

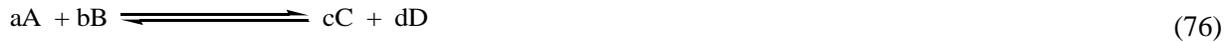

aA + bB ⇌ cC + dD  (76)

Equilibrium constant for this reaction in terms of partition function for moving observer given is defined as,

$$(K_{eq})_u = [Q_C]^c [Q_D]^d / [Q_A]^a [Q_B]^b \exp(-\langle \Delta \varepsilon_0 \rangle_u / R_u T) \qquad (77)$$

Placing value of Lorentz transformations of difference in zero point energies of reactants and products and ideal gas constant i.e. $\langle \Delta \varepsilon_0 \rangle_u = \gamma^{-1} \langle \Delta \varepsilon_0 \rangle_o$ and $R_u = \gamma^{-1} R_0$ in Eq. (77) proves Equilibrium constant to be Lorentz invariant,

$$(K_{eq})_u = (K_{eq})_o = K_{eq} \qquad (78)$$

System at chemical equilibrium should appear the same for observers at rest and in motion. No doubt rate of forward and backward reaction slows down as both rate and rate constant of forward and backward reactions are Lorentz variant. Equilibrium constant being ratio of rate constants for forward and backward reactions remains Lorentz invariant. As total molecular partition function is Lorentz invariant so this renders Equilibrium constant also Lorentz invariant. Thus amount of reactant and product in equilibrium with one another remain the same for all observers independent of their relative speeds.

## 8. Discussions

### 8.1. Discussion on relativistic statistical thermodynamics

Present theory of relativistic statistical thermodynamics is unification of special theory of relativity with statistical thermodynamics. Since special theory of relativity is based on two basic postulates that all laws of physics retain their form in all inertial frames. So Avramov's view that Lorentz transformation of Boltzmann constant is a tempting problem for statistical thermodynamics contradicts with the second postulate of special relativity. Present theory of relativistic statistical mechanics shows that Lorentz transformation of Boltzmann constant is consistent with the second postulate of special relativity and does not put any serious problems on statistical physics. Relative life time of molecular transition between two states depends upon the spacing between two energy levels. Greater the spacing between two allowed energy states quicker the molecule de-excites and vice versa. Life time of transition is inversely related to spacing between the two levels that is why electronic transitions are quicker than vibronic transitions and vibronic transitions are greater than rotational transitions of molecules. Same indirect proportionality between life time and spacing between quantum levels is utilized in lasers to achieve population inversion. As it follows from special theory of relativity that time period of an event stretches for moving observer so to be consistent with nature life time of an excited state also increases which occurs on behalf of decrease in energy spacing between permitted quantum states this makes energy spacing Lorentz variant i.e. $\langle \Delta \varepsilon_u \rangle = \gamma^{-1} \langle \Delta \varepsilon_o \rangle$. From spectroscopic studies of isotopomers it has been found that when lighter isotopes in molecules are substituted with one of its heavier isotope zero point energies of molecules decreases and spacing between permitted rotational and vibrational level also decreases. In nature it is found that increase in mass of molecule from isotopic substitution results in lowering energy of each and every allowed quantum level associated with molecule thus energy spacing between successive quantum levels decreases.[33] As it follows from special theory of relativity mass increases for moving observer i.e. $m_u = \gamma m_o$, so Lorentz transformation of energy spacing between permitted quantum levels is compatible with Lorentz transformation of mass. Thus increase in mass of molecule resulting in decrease in spacing of permitted quantum levels for moving observer is analogous to lowering of rotational and vibrational levels in heavy isotopomers.[34] Mathematical equations describing distribution of total number of particles and total amount of energy in system among permissible energy levels possess an exponential factor i.e. $w_o = exp(\langle -\Delta \varepsilon_o \rangle/(k_B)_0 T)$. Since both spacing between successive energy levels and Boltzmann constant are Lorentz variant i.e. $\langle \Delta \varepsilon_u \rangle = \gamma^{-1} \langle \Delta \varepsilon_o \rangle$ and $(k_B)_u T = \gamma^{-1}(k_B)_0 T$ respectively, so this renders exponential factor Lorentz invariant i.e. $w_u = w_o = w$ which is common to both partition function and distribution of molecules among their permitted energy states. So number of molecules in particular energy level will remain same for all observers independent of their relative speeds. Thus population distribution among quantum levels is identical for observer at rest and observer moving at fractions of speed of light. To prepare system with total energy $E_u$ one has to distribute it among total number of molecules and among all internal degrees of freedom (translational, rotational, vibrational and electronic) of these molecules. Distribution of $E_o$ among total number of molecules and among all internal degrees of freedom will decrease for moving observer but ratio with which it is distributed among total number of molecules and among all internal degrees of freedom will remain same for all observers independent of their relative speeds. So ratio of distribution of energy among total number of molecules and among their all internal degrees of freedom is same for observers irrespective of their speeds. So total molecular partition function can be factorized into contribution from each form of energy i.e. translational, rotational, vibrational and electronic is also Lorentz invariant. Thermal de Broglie wavelength associated with molecules i.e. $\Lambda_u = \sqrt{2\pi m_u (k_B)T}/h$ can also be mathematically proved to be Lorentz invariant by substituting Lorentz transformation of mass and Boltzmann constant i.e. $\Lambda_u = \Lambda_o = \Lambda$ Critical temperature which is defined as temperature at which the thermal de Broglie wavelength is on the order of, or larger than the interparticle distance so that quantum effects will dominate and the gas must be treated as a Fermi gas or a Bose Gas, obeying Fermi-Dirac or Bose Einstein Statistics. Since according to Landsberg and Matsas there exists no law of temperature transformations under Lorentz boosts and it remains the same for all observers independent of their relative speeds. As explained by Avramov if the system is at a triple point of the corresponding *P–T* diagram three phases will be visible for all observers irrespective of their relative speeds. As soon as they can coexist at one *P*, *T* point only, the temperature and the pressure of this system will be same for observers in rest and moving frames, on the same grounds critical temperature at which Maxwell Boltzmann statistics switches to Fermi-Dirac or Bose-Einstein statistics should be the same for all observers independent of their relative speeds. This also support Thermal de Broglie wavelength to be Lorentz invariant. Thus increase in mass of molecules with decrease in spacing of energy levels renders translational partition function to be Lorentz invariant. Ignorance of Lorentz transformation of length in translational partition function is very much similar to that is done for concentration of solutions. Volume contracts for moving observer but concentration of solutes (ions, molecular species) remain invariant for moving observer because Lorentz transformation of length is being ignored for concentration of solutions and so same treatment is practiced for translational partition function. This point can be explained by considering particle in a symmetric box like cubic box, in which existence of symmetry gives birth to many degenerate energy levels. So if Lorentz transformation of length is taken in account it will break the symmetry of box and will uplift the degeneracy among the different energy states. This would lead degenerate energy levels to become non-degenerate for moving observer which leads to an absurd situation. To prohibit symmetry violation and uplifting of degeneracy among different energy states for moving observer transformation of length of box is ignored and treated Lorentz invariant for thermal partition function. No doubt energy of degenerate energy levels should decrease for moving observer but their degeneracy remains Lorentz invariant for moving observer.

As translational temperature is Lorentz invariant so extending same concept to rotational, vibrational and electronic temperature they all have been mathematically proved invariant under Lorentz boosts. For moving observer mass of molecules increases so this result in increase in moment of inertia of molecules thus making it Lorentz variant i.e. $I_u = \gamma I_o$. According to law of conservation for rotating body product of moment of inertia and angular velocity is constant, so for moving observer rotational speed of molecules slows down i.e. $\omega_u = \gamma^{-1} \omega_o$ due to Lorentz transformation of moment of inertia i.e. $I_u = \gamma I_o$. This lowering of rotational speed of molecules is totally in agreement with phenomenon of time dilation which is most important experimentally verified consequence of special relativity. [6-7] Lorentz transformation of bond will not be considered as rotating molecules do not align along transformed axis. Moreover for moving observer mass of electron increase so according to Heisenberg's Uncertainty principle i.e. "$m \Delta v \Delta x \geq \hbar$" velocity of electron revolving slows down while space occupied by electrons remains invariant i.e. why bond length is Lorentz invariant. During rotation of molecules bond length will not be along the direction of motion of moving observer so it will not suffer length contraction. Increase in moment of inertia of molecules lowers rotational energy levels and thus rotational constant decreases $B_u = \gamma^{-1} B_o$ for moving observer. This is again in agreement with time dilation as moving observer will observe molecules rotating slowly and time period for their one complete round trip stretches or increases. Relativistic rotational constant and relativistic Boltzmann constant all together renders rotational temperature Lorentz invariant. This makes rotational partition function Lorentz invariant as shown. Molecules are oscillators so their atoms do vibrate, as mass of atoms increases for moving observer so this lowers their vibrational frequency making it Lorentz variant i.e. $v_u = \gamma^{-1} v_o$, this is also totally in agreement with time dilation as in moving frame time dilates, so time for one complete vibration will also stretch and atoms will slowly execute compression and extension thus vibrational frequency lowers down. Lowering of vibrational frequency and Boltzmann constant for moving observer all together makes vibrational temperature Lorentz invariant. Vibrational partition function which depends on vibrational temperature also becomes Lorentz invariant. Lowering of rotational and vibrational energy levels in moving frame is analogous to substitution of heavier isotopes in place of lighter ones in molecules. On the same grounds electronic temperature and electronic partition function is proved Lorentz invariant. Thus total Molecular partition function becomes Lorentz invariant.

All thermodynamic state variables have been proved Lorentz variant in relativistic statistical thermodynamics. All thermodynamic state variables in some way or the other are described in terms of thermodynamic work associated with the system. Since thermodynamic work is Lorentz variant and appears to decrease for moving observer so all thermodynamic state variables decrease for moving observer. [18-19] So Lorentz transformational of all state functions is compatible with Lorentz transformation of thermodynamic work. Decrease in entropy is also due to Lorentz transformation of Boltzmann constant. In accordance with time dilation moving observer should find universe younger than observer in rest frame. Since it is known from second law of thermodynamics that universe is expanding since it's time of creation and accompanied with increase in entropy. [35] So age of universe is directly related to entropy of universe. To be in accord with time dilation moving observer should observe universe younger with it's to entropy be than that observed by observer at rest. So Lorentz transformation of entropy is in accord with time dilation. This explanation is similar to that of twin paradox. [3]

## 8.2. Discussion on relativistic chemical kinetics

Transition state theory which is the most general and universal theory of chemical kinetics is also very successful in evaluating absolute reaction rates. This theory introduces concept of activated complex called transition state whose formation is responsible for conversion of reactants in to products by executing translational motion along the reaction coordinate. According to Eq. (2) mass of transition state increases for moving observer i.e. $m_u^\dagger = \gamma m_o^\dagger$ while according to Eq. (59) velocity of transition state decreases for moving observer i.e. $v_u^\dagger = \gamma v_o^\dagger$ so this renders momentum of transition state to be Lorentz invariant i.e. $\Delta p_u = \Delta p_o = \Delta p$. It is known from de Broglie relation that mass and de Broglie wavelength associated with a transition state are inversely related to one another i.e. $\lambda_u = h/\Delta p_u$.[36] As momentum of transition state is Lorentz invariant i.e. $\Delta p_u = \Delta p_o = \Delta p$, so de Broglie associated with transition state is independent of relative speeds of moving observers i.e. $\lambda_u = \lambda_o = \lambda$. Since for moving observer transition state is more massive than transition state for observer at rest so for different observers to agree on de Broglie wavelength they should disagree on velocity of transition state along the reaction coordinate, this is in agreement with phenomenon of time dilation. Moving observer should observe less velocity of transition state along the reaction coordinate than that observed by observer at rest. According to Heisenberg's Uncertainty principle reaction coordinate must be at least the size of de Broglie wavelength associated with transition state i.e. "$\Delta q_u = \lambda_u/2\pi$".[25] So de Broglie wavelength associated with transition state is Lorentz invariant i.e. $\lambda_u = \lambda_o = \lambda$, this renders reaction coordinate to be Lorentz invariant i.e. $\Delta q_u = \Delta q_o = \Delta q$. Concentration of activated complexes in a length along the reaction coordinate of length $\delta_u^\dagger = h/(2\pi m_u^\dagger (k_B)_u T)^{1/2}$ becomes Lorentz invariant on substitution of Lorentz transformations for Boltzmann constant and mass i.e. $\delta_u^\dagger = \delta_o^\dagger = \delta^\dagger$. The average rate of passage of activated complexes over the reaction barrier in one direction along the coordinate of decomposition is $r_u^\dagger = ((k_B)_u T / 2\pi m_u^\dagger)^{1/2}$ which becomes Lorentz variant on substitution of Lorentz transformations for Boltzmann constant and mass i.e. $r_u = \gamma^{-1} r_o$. Therefore, for moving observer transition state possesses less kinetic energy and executes slow translational motion on the reaction coordinate. Lorentz transformation of the Boltzmann constant deprives the transition state with kinetic energy associated with it. Therefore, Lorentz transformation of the Boltzmann constant supports the time dilation for motion of activated complexes along the reaction coordinate. Rate constant of fastest reaction for moving observer is equal to "$(k_B)_u T / \hbar$".[17] Therefore, decrease in Boltzmann constant is responsible for the decrease in rate constant as frequency for the passage through the transition state slows down. This is in accordance with the special theory of relativity according to which time dilates, as frequency is the reciprocal of time so frequency should be observed to decrease for moving observer and time period of reaction stretches. Lorentz transformation of rate constant equation is applicable to all kinds of reactions regardless of what type of kinetics they follow i.e. zero order, first order, second order and third order etc. Transitions state theory gives thermodynamic definition of Arrhenius factor. Since all thermodynamic state functions and universal gas constant are Lorentz variant, this makes Arrhenius factor Lorentz variant as shown in Eq. (53). Arrhenius factor is pre-exponential factor in rate equation and contributes towards frequency for the passage through the transition state as mass of reacting atoms and molecules increases for moving observer so frequency is observed to decrease and thus Arrhenius factor is observed to decrease for moving observer which is totally in agreement with time dilation phenomenon. Rate constant gives quantitative knowledge about the speed of reaction and as the rate constant decreases the rates of reaction should also slow down. This is mathematically proved for the first time that rate of reaction is Lorentz variant as shown in Eq. (73). Lorentz transformation of the half- life period equation is derived first time here as shown in Eq. (75) and it is precisely similar to Einstein's time dilation equation as shown in Eq. (1). This strongly supports the present theory of rates of reactions to be consistent with special theory of relativity. Slowing down the rate of reaction in moving frame is more or less analogous to kinetic isotopic effect in chemistry. [37] When heavier isotopes are present in molecules rate of reaction slows down. Collision theory explains kinetics of bimolecular reactions and expresses rate of reactions as frequency of bimolecular collisions occurring in reacting molecules. Rate of reaction depends upon number of fruitful collisions occurring per second. Frequency of collision is responsible for rate of reaction. Rate constant in collision theory is defined by product of relative velocity of colliding molecules and collision cross section area involved in the bimolecular collision. Increase in mass of molecules lowers the velocity of molecules so relative velocity between the colliding molecules observes to decrease for moving observer. Decrease in Boltzmann constant also lowers the energy available per molecule, this decrease in energy per molecule is equivalent to increase in mass per molecule, thus slows down the relative velocity between the molecules. Collision frequency which is related to rate of reaction is product of area of collision cross section, mole densities and relative velocity of colliding molecules. Since collision cross section area and mole densities (concentration) are Lorentz invariant, while relative velocity is Lorentz variant thus relative velocity renders collision frequency to be Lorentz variant as shown in Eq. (63). Further, collision frequency decreases for moving observer and thus rate also slows down this is in accordance with relativistic rate equation. Relativistic collision frequency equation is in accordance with special theory of relativity as frequency is the reciprocal of time and as time dilates, so does frequency and it is observed to decrease for moving observer. The Marcus theory is statistical mechanical and employs actual potential energy surfaces and actual atomic coordinates to describe a number of important processes in chemistry and biology, including photosynthesis, corrosion, certain types of chemiluminescence's charge separation in some types of solar cell and more. [30-32] Besides the inner and outer sphere applications, Marcus theory has been extended to address heterogeneous electron transfer. Electron transfer occurs from donor to acceptors. These transfers occur much faster than nuclear vibrations. Therefore, the nuclei do not appreciably change their position during the time of electron transfer. During the transfer, the electron does not change energy i.e. the energy of the donor and acceptor orbital must be the same prior to transfer. The energy levels of the donor and acceptor orbitals in the reactants and products are in continual flux due to internal nuclear movements and the solvent motions. For transfer, the donor and acceptor molecules must simultaneously achieve particular geometries and solvation arrangements that give matched energy levels between the donor and acceptor orbitals. After electron transfer, the nuclei of donor and acceptor molecules relax to their optimum positions. The energy required to change the solvation sphere and internal structures bringing the donor and acceptor orbitals to same energy is called the reorganization energy. This energy creates barrier to electron transfer. Reorganization energy is defined as energy that needs to distort either the reactant-solvent or product-solvent ensemble, or into both, to make the energy of the donor and acceptor orbitals the same.

Vibrational reorganization energy $(\lambda_i)_0$ measures the energy difference due to changes in bond length, angles etc. which occur upon electron transfer as shown in Eq. (65). Increase in mass of nuclei of reacting species decreases their characteristic oscillation frequency and renders it as Lorentz variant, and thus reduced force constant is observed to decrease for moving observer. So energy require due to changes in bond length, angles etc. for successive transfer of electron decreases for moving observer would be less for successful electron transfer as it is shown in Eq. (66). While solvational reorganization energy $(\lambda_0)_o$ measures the energy involved in reorganization of the solvent shell for electron transfer as shown in Eq. (66). When electron transfer reaction is carried out in moving frame increase in mass for moving observer makes density Lorentz variant. Refractive index is directly proportional to density greater the density of material greater will be its refractive index. Therefore, for moving observer solvent becomes denser than for stationary observer and hence refractive index and dielectric constant become Lorentz variant thus rendering less solvational reorganization energy required for successive transfer of electron which is shown in Eq. (67).This results in Lorentz transformation of total reorganization energy for moving observer. In electron transfer reaction, there is a very little spatial overlap of the electronic orbitals of the two reacting molecules in activated complex. The assumption of slight overlap leads to an intermediate state X[*] in which electric polarization of the solvent does not have the usual value appropriate for the given ionic charges. The intermediate state X[*] can either disappear to reform reactants, or by electronic jump mechanism to form a state X in which the ions are characteristic of products. Rate constant equation for such electron transfer reaction possess electronic coupling $(H_{AB})_o$ as pre-exponential factor as which is due to assumption that there is slight spatial overlap of $X^*$ and X and thus energies of two states are equal. The energy of any state, $X^*$ is broadened by amount $(\Delta\varepsilon)_o$ which is related to life time $(\Delta t)_o$ of the state according to Heisenberg Uncertainty principle "$\langle\Delta\varepsilon_o\rangle\langle\Delta t_o\rangle \geq \hbar$". *The greater the overlap shorter will be the life times of states $X^*$ and X.* [36] For moving observer electronic coupling between the $X^*$ and X decreases because in moving frame mass of electron increase so according to Heisenberg's Uncertainty principle i.e. "$m\Delta v\Delta x \geq \hbar$" velocity of electron revolving slows down while space occupied by electrons remains invariant i.e. why bond length is Lorentz invariant. Slower movement of electrons results in the decrease of coupling between interacting electronic clouds and thus electronic coupling becomes Lorentz variant i.e. $(H_{AB})_u = \gamma^{-1}(H_{AB})_o$. According to time dilation, the life time of states $X^*$ and X will increase i.e. $\langle\Delta t_u\rangle = \gamma\langle\Delta t_o\rangle$ in moving frame so this will result in smaller overlap of orbitals and thus electronic coupling is observed to decrease in moving frame and energy of states is given by $\langle\Delta\varepsilon_u\rangle = \gamma^{-1}\langle\Delta\varepsilon_o\rangle$. For moving observer the energy of any state, $X^*$ is broadened by amount $\langle\Delta\varepsilon_u\rangle$ which is related to life time $\langle\Delta t_u\rangle$ of the state according to Heisenberg Uncertainty "$\langle\Delta\varepsilon_u\rangle\langle\Delta t_u\rangle = \gamma\langle\Delta t_o\rangle\gamma^{-1}\langle\Delta\varepsilon_o\rangle \geq \hbar$". *The smaller the overlap greater will be the life times of states $X^*$ and X.* Heisenberg Uncertainty principle remains valid in moving frame. It is consistent with basic postulate of special theory of relativity that laws of physics holds valid in all inertial frames independent of their relative speeds. Therefore, rate constant for electron transfer reaction is consistent with relativistic rate constant equation and thus the rate of reaction also slows down.

## 9. Applications

### 9.1. Enzymes Kinetics

Methylamine dehydrogenase (MADH) is a quinoprotein that converts primary amines to aldehyde and ammonia. Consider a reaction shown in Fig.1 in which methyl amine is converted to formaldehyde. [38]

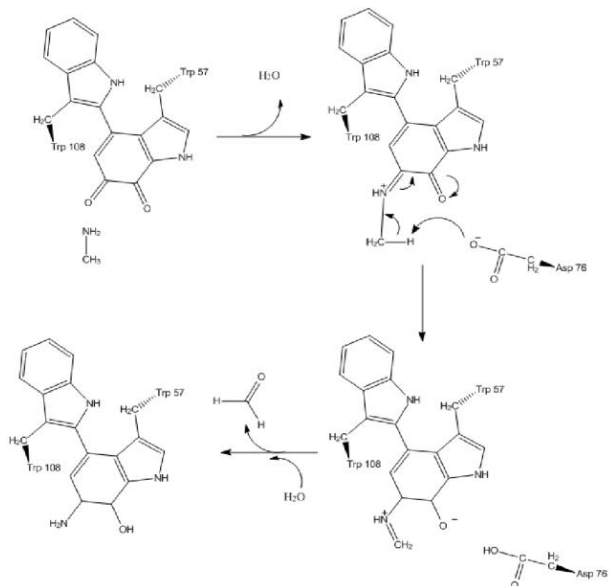

Fig.1 Methyl Amine oxidizing methyl amine obeying first order kinetics

Methylamine dehydrogenase obeys first order kinetics. Key step in this enzymatic reaction is hydrogen transfer which also occurs due to tunneling effect. Let methylamine dehydrogenase and methyl amine have unit concentrations so that value of rate constant and rate of reaction are equal. [27] Now, if the same enzymatic reaction is carried out in moving frame at a speed of $2\times10^8$ ms$^{-1}$ or moving observer monitors reaction at speed of $2\times10^8$ ms$^{-1}$ then rate of enzymatic reactions slows down and rate constant is also observed to decrease as shown in the table 1.

| Table 1. Comparison of rate, rate constant and half- life (in moldm$^{-3}$s$^{-1}$ and sec respectively) of Proton Transfer Reaction in Methylamine Dehydrogenase at 298 K for moving and stationary observer. | | | | | |
|---|---|---|---|---|---|
| Rate of reaction for stationary observer [a] | Rate constant of reaction for stationary observer [a] | Half life of reaction of for stationary observer [a] | Rate of reaction for moving observer [b] | Rate constant of reaction for moving observer [b] | Half life of reaction of for stationary observer [b]] |
| 0.0087 | 0.0087 | 79.6 | 0.0064 | 0.0064 | 106.7 |
| [a] This data has been developed from Eq.s (73) and (75)  [b] This date has been developed from Eq.s (52),(74) and (76) | | | | | |

## 9.2. Radioactive disintegration of Astatine

Astatine $^{211}$At$_{85}$ emits alpha particles and decays in to Bismuth $^{207}$Bi$_{83}$ having a half life of 7.2 hours. It is used in radio immunotherapy. [39-40]

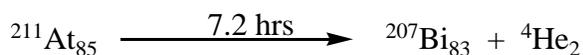

$^{211}$At$_{85}$ $\xrightarrow{\text{7.2 hrs}}$ $^{207}$Bi$_{83}$ + $^{4}$He$_{2}$

Now if the same nuclear reaction of one mole of astatine is carried out in moving frame moving at a speed of $2\times10^8$ m/s or observer moving at a speed of $2\times10^8$ m/s monitor same nuclear reaction then rate of decay of astatine is observed to slow down and half life is dilated as shown in the table 2.

| Table 2. Comparison of rate, rate constant and half- life (in moldm$^{-3}$s$^{-1}$ and hours respectively) of radioactive decay of Astatine $^{211}$At$_{85}$ in to Bismuth $^{207}$Bi$_{83}$ for moving and stationary observer. | | | | | |
|---|---|---|---|---|---|
| Rate of reaction for stationary observer [a] | Rate constant of reaction for stationary observer [a] | Half life of reaction of for stationary observer [a] | Rate of reaction for moving observer [b] | Rate constant of reaction for moving observer [b] | Half life of reaction of for stationary observer [b]] |
| 2.67×10$^{-5}$ | 2.67×10$^{-5}$ | 7.2 | 1.93×10$^{-5}$ | 1.93×10$^{-5}$ | 9.6 |
| [a] This data has been developed from Eq.s (73) and (75)  [b] This date has been developed from Eq.s (52),(74) and  (76) | | | | | |

The emission of alpha particles from radioactive elements involves tunneling of the alpha particles through the potential energy barrier produced by the short range attractive nuclear forces and the coulombic repulsive force between the daughter nucleus and the alpha particles. Now when reaction is carried out in moving frame moving at a speed of $2\times10^8$ m/s or observer moving at a speed of $2\times10^8$ m/s then increase in mass of neutrons and protons is observed in the nucleus and so mass of alpha particles is also observed to increase and thus it becomes difficult for heavier alpha particles to tunnel through the barrier generated by short range attractive nuclear forces and the columbic repulsive force between the daughter nucleus and the alpha particles as compared to lighter alpha particles in rest frame. So slowing down of tunneling slows down the rate of radioactive decay and thus half life of astatine dilates. Since it is known that radioactivity can't be slowed down by either lowering or increasing the temperature. So time dilation of half life of radioactive elements does not support either of any Lorentz transformation of temperature and also confirms temperature to be Lorentz invariant. Slowing down of radioactive process for moving observer is due to Lorentz transformation of its rate constant. Lorentz transformation of radioactive rate constant strongly supports Lorentz transformation of Boltzmann constant.

## 10. Conclusions

In quantum theory energy bears the same relation to time as space does to time in theory of special relativity. So nature has knotted time and energy in a same manner as it has knotted space and time. That is why time dilation dictates energy spacing between permitted quantum levels to be Lorentz variant. So product of time and energy in uncertainty relation is Lorentz covariant. For moving observer ratio of energy spacing between permitted quantum levels and Boltzmann constant is Lorentz invariant and thus population of permitted energy levels remains unchanged. Time dilation at molecular level is possible due to Lorentz transformations of Boltzmann constant, energy of permitted quantum states and all thermodynamic state functions. As magnitude of rate of reaction depends upon relative speeds of observer, therefore, it's not absolute just like space and time. As rate of reaction is dependent on relative speeds of moving observer, therefore presented theory rejects the concept of absolute reactions rate.

### Acknowledgements

The author greatly acknowledges Czech academy of sciences for financial support.